\documentclass[aps,prb,showpacs,groupedaddress,twocolumn]{revtex4}
\usepackage{bm}
\usepackage{graphicx}
\newcommand{\Efield}{{\mathcal{E}}}

\newcommand{\bes}{\begin{eqnarray}}
\newcommand{\ees}{\end{eqnarray}}
\newcommand{\be}{\begin{equation}}
\newcommand{\ee}{\end{equation}}
\begin{document}
\title{Kinetic theory of quantum transport at the nanoscale}
\author{Ralph Gebauer}
\affiliation{The Abdus Salam International Centre for Theoretical
Physics (ICTP), 34014 Trieste, Italy \\
INFM Democritos National Simulation Center, 34013 Trieste, Italy}  
\author{Roberto Car}
\affiliation{Department of Chemistry and Princeton Institute for the
  Science and Technology of Materials, Princeton University,
  Princeton, New Jersey 08540 }  
\date{\today} 
\begin{abstract}   
We present a quantum-kinetic scheme for the calculation of
non-equilibrium transport properties in nanoscale systems. The
approach is based on a Liouville-master equation for a reduced density
operator and represents a generalization of the well-known Boltzmann
kinetic equation. The system, subject to an external
electromotive force, is described using periodic boundary conditions. 
We demonstrate the feasibility of the approach by applying it to a double-barrier
resonant tunneling structure.
\end{abstract}
\maketitle

\section{Introduction}

There is currently large interest in the development of electronic devices 
that operate at the molecular level, since these may revolutionize the electronic industry 
\cite{Ratner}. Modeling electron transport at the nanoscale
poses great theoretical challenges because in this regime the semiclassical Boltzmann
kinetic equation commonly used to describe transport is not applicable.
Boltzmann's equation treats the electrons in terms of their classical probability distribution in
phase space, and deals with collision events quantum-mechanically
using Fermi's golden rule \cite{Rammer}. This formalism explains why 
conduction electrons subject to a uniform electric field do not accelerate indefinitely, but
settle in a steady state regime of constant current, as a consequence of inelastic collisions 
with the lattice that lead to energy dissipation.

To describe transport in devices having spatial dimensions of the
order of the electron wavelength,  
Boltzmann's equation should be replaced by a quantum mechanical
Liouville-master equation for a reduced density operator
\cite{Rammer, FischettiMaster, FischettiPauli, RossiPRL, Rossi}. 
Such quantum kinetic approaches to electron transport have to deal
with several difficulties. Probably most severe are those related to
the description of boundary conditions between the device active
region and the contacts: In most common approaches
electrons can be exchanged between the device and its environment
through  left ($L$) and right ($R$) contacts with the electrodes of a
battery. The $L$($R$) injected electrons are assumed to have thermal 
equilibrium distributions defined by the temperature $T$  and the 
electrochemical potentials $\mu_{L(R)}$ of the electrodes. 
In self-consistent calculations, however, modifications of the 
distribution of the injected carriers, like drifting the momentum of 
the distribution or displacing the chemical potentials of the
contacts, are necessary to avoid unphysical charging 
of the contacts \cite{Poetzqw}. Overall the issue of contacts and 
boundary conditions is crucial in all models of quantum 
transport \cite{Rossi, Laux, Poetzqw, Frensley,  PoetzDB, Vashishta, 
Lent}.

In this paper we present a kinetic formulation which avoids these
difficulties by adopting periodic boundary conditions. The active
region together with a part of the contacts is placed in a unit cell
that is repeated periodically. One can imagine such a geometry as an
infinite ring, without open boundaries. In this way we deal with a
closed circuit with no exchanges of electrons with the environment
\cite{KohnLutt, BuettikerPRL, BIL}. A current is induced in the
circuit by an externally applied electromotive force.

In our approach coupling between the electrons and the lattice is
included explicitly to prevent the electrons from accelerating
indefinitely. This coupling is treated in perturbation theory (weak coupling). 
Furthermore, the time scale of inelastic
electron-phonon collisions is typically much faster than the current
relaxation time and is eliminated by a 
{\it coarse graining} procedure. 
The resulting Markovian dynamics of the reduced electron density
operator is described by a Liouville-master equation,
which effectively generalizes Boltzmann's equation at the
nanoscale \cite{FischettiPauli, RossiPRL}. 
We treat the diagonal and off-diagonal elements of the density
operator on equal footing.  

One common problem of kinetic master equations that describe
collisions using Fermi's golden rule is that current continuity  is
violated in inhomogeneous media when the current density is defined in
the standard way \cite{FischettiPauli, Frensley}. In our approach,  
continuity is exactly satisfied, by properly taking into account the
current contribution due to electron-phonon collisions, as described
elsewhere \cite{currentOS}.
 
The reminder of this paper is organized as follows. In section
\ref{methods} we describe the methods that we use. In particular,
we recall how a Liouville master equation for an electron 
in contact with an external heat bath is derived (section \ref{lme}). 
We then discuss electric fields (section
\ref{efield}), the definition of currents (section \ref{currents}),
and a generalization of the formalism to many particle systems
(\ref{manyp}). Finally, in section \ref{example}, 
we demonstrate the scheme with a numerical
application to a system consisting of a one-dimensional wire
connected to a double-barrier resonant tunneling structure
(DBRTS). This example illustrates the key features of our approach and
the differences between our treatment and open boundary formulations
\cite{PoetzDB,Poetzqw, Frensley, Buettikerbarriers}.  

\section{Method}
\label{methods}
\subsection{Liouville master equation}
\label{lme}

For simplicity we use in the following a one-dimensional notation, but
extension to 
three-dimensions is straightforward. A ring of length $L$ is a
simple periodic structure. Ideally we should consider an infinite ring
($L \rightarrow \infty$). In numerical applications, however, $L$ must be finite.
This could
be achieved by placing many periodically repeated unit cells of length
$L$ in a ring, but in the following we shall consider only the case of a
single periodic unit having length equal to the perimeter of the
ring. In absence of the
external electric field the single-electron Hamiltonian is 
$H_0 = 1/2 p^2 + U(x) = \sum_n \varepsilon_n c_n^\dag c_n$, where $U(x)$ is a
potential with periodicity $L$, the sum is over eigenstates of energy $\varepsilon_n$,
and $c_n^\dag (c_n)$ are electron creation (annihilation) operators. Atomic units 
($e = \hbar = m = 1$) are adopted throughout. The harmonic
phonon bath has Hamiltonian $R = \sum_{\alpha} \omega_{\alpha} a^\dag_{\alpha}
a_{\alpha}$, 
where the sum is over eigenmodes of
frequency $\omega_{\alpha}$ , and  $a^\dag_\alpha (a_\alpha)$ are phonon 
creation (annihilation) operators. The
electron-phonon coupling potential is 
$V = \sum_{m,n,\alpha} \gamma^\alpha_{m,n} c^\dag_n c_m a^\dag_\alpha + h.c.$, 
in terms of matrix elements $\gamma^\alpha_{m,n}$ ($h.c.$ denotes the hermitian
conjugate).
The Hamiltonian of the total system (electron + bath) is $H_T = H_0 + R + V$. The
density operator for the total system  $\Sigma$ satisfies the Liouville
equation $\frac{d \Sigma}{dt} = -i \left[H_T, \Sigma\right]$. The reduced 
density operator for the electronic subsystem, $S$,
is obtained from $\Sigma$ by tracing out the bath degrees of freedom,
{\it i.e.} $S = \mbox{Tr}_R \Sigma$. The equation of motion for $S$ 
can be derived using a standard
procedure \cite{Louisell,Cohen-Tannoudji} in which $V$  is treated to second order in perturbation
theory, and a Markov approximation is made, whereby  electron-phonon
collisions become instantaneous processes described by Fermi's golden
rule. The bath is assumed to be in thermal equilibrium at all times,
i.e. any retroaction of the electron system on the bath is
neglected. Then the Liouville-master equation for $S$ is \cite{Louisell,Cohen-Tannoudji}:  
\be
\label{master}
\frac{dS}{dt} = -i\left[H_0,S\right] + {\cal L}\left[S\right],
\ee

Here the term that describes collisions with the bath is \bes
{\cal L}\left[S\right] \equiv -\sum_{m,n} && \Gamma_{mn} \big(
L_{nm}L_{mn}S + \nonumber \\
&& S L_{nm}L_{mn} - 2L_{mn}SL_{nm} \big),
\ees
where $L_{nm} = c^\dag_n c_m$, and the transition probabilities
$\Gamma_{mn}$ are given by:
\bes
\label{Gamma}
\Gamma_{mn} = \left\{
\begin{array}{ll}
{\cal D}(\varepsilon_n - \varepsilon_m) \left|\gamma_{mn}\right|^2
\left(\bar{n}_{\varepsilon_n-\varepsilon_m} + 1\right) & \varepsilon_n >
\varepsilon_m \\
{\cal D}(\varepsilon_m - \varepsilon_n) \left|\gamma_{mn}\right|^2
\bar{n}_{\varepsilon_m-\varepsilon_n}  & \varepsilon_m > \varepsilon_n. 
\end{array}\right.
\ees

Here $\bar{n}_\omega$ is the thermal occupation and ${\cal D}(\omega)$ is the density
of states of phonons with frequency $\omega$. Eq.~(\ref{Gamma}) ensures detailed 
balance leading the electron subsystem to thermal
equilibrium in absence of an applied external electric field. 

In our scheme, we calculate the time evolution of the 
density matrix, including diagonal and off-diagonal elements. In
approaches using open boundary conditions, it is common practice
to  limit the description to the diagonal elements of the density matrix
only \cite{FischettiPauli}. It should be stressed, however, that the representation
of the density matrix is very different when open boundary conditions are used
instead of periodic boundary conditions. In open boundary approaches the different chemical 
potentials of the $L$ and $R$ contacts cause a beaking of symmetry between the
contacts, and the final state is largely determined by 
the different occupations of the $L$ and $R$ incident channels.  
These occupations are given by the diagonal elements of the density matrix
in the open boundary representation. In our approach, where
periodic boundary conditions are adopted, we need to use {\it unperturbed eigenfunctions} as basis states. 
These states do not carry current, and the final, current-carrying state
of the system is dominated by the off-diagonal elements of the density
matrix in this basis. The computational complexity remains limited, however, 
because not all the elements of the density matrix need to be considered, but only those that 
fall within a finite energy window around the Fermi energy.

\subsection{External static electric field}
\label{efield}

We include an external uniform static electric field $\Efield$ by letting
$p \rightarrow p - 1/c A(t)$.
The Hamiltonian becomes $H(t)=1/2 \left(p - 1/c A(t)\right)^2 + U(x)$. 
The vector potential is $A(t) = -c \Efield t$. This choice of the
gauge ({\it v-gauge}) is compatible with periodic boundary conditions.
We avoid the difficulty with a Hamiltonian that grows indefinitely
with time by systematically performing gauge transformations in which 
the electronic wavefunctions are multiplied by a phase factor $\exp(-i
A(t)/c \cdot x)$ and $A(t)$ in the Hamiltonian is restored to its initial value $A(t=0)$
\cite{KohnIns}. In this way
all the time dependence is transferred to $S$ and the Hamiltonian remains constant.
In order not to violate periodic boundary conditions  
the gauge transformations are only possible at times that are integer multiple
of $\tau_\Efield = \frac{2\pi}{L \Efield}$. 
The time $\tau_\Efield$ is arbitrarily
small for $L \rightarrow \infty$, but is always finite in numerical
applications. In practice we proceed as follows. Let $A(0)=0$
at $t=0$. First we integrate Eq.~(\ref{master}) 
for the time period $\tau_\Efield$ using the Hamiltonian
propagator with the electric field, which corresponds to the first
term in the right hand side (r.h.s.) of Eq.~(\ref{master}). 
Then we apply a gauge transformation to set 
the vector potential to zero. Finally, we propagate $S$ for
the same period $\tau_\Efield$ using the collision 
propagator, which corresponds to the second term on the r.h.s. of
Eq.(\ref{master}). The result is a density matrix
$S(\tau_\Efield)$. This procedure is repeated at all subsequent time
steps in increments of $\tau_\Efield$. In the limit of $\tau_\Efield
\rightarrow 0$ this procedure converges to the exact kinetic
propagation. 

Our approach in which a static external electric field acts as driving
force for the electrons, is well suited to dealing with systems where the voltage drop
occurs along one given direction in space. Complex
geometries in devices with many-terminals \cite{Laux} may not 
be tractable in this framework. However, many current
applications of nano-scale transport, like {\em e.g.} break
junctions, STM tips, or interlayer tunneling, can be represented
within our scheme. 

\subsection{Hamiltonian and dissipative currents}
\label{currents}

The current density associated to Hamiltonian propagation is 
\bes
\label{current}
j(x;t) &=& \mbox{Tr}\left[ S(t) \hat{J}(x;t) \right], \nonumber \\
\hat{J}(x;t) &=& \frac1{2} \Big[ \big(\hat{p}-1/c A(t)\big)
  \delta(x-\hat{x}) \nonumber \\
&&+ \delta(x-\hat{x})\big(\hat{p}-1/c A(t)\big)
  \Big],
\ees
where the current (as well as all other observables) is averaged over
the period $\tau_\Efield$.
It is important to note that these definitions do not yield a
current that satisfies the continuity equation
\be
\label{cont}
\left<x\left| \frac{d S(t)}{dt} \right| x \right> + \frac{d}{dx}
j(x;t) = 0.
\ee
In contrast, Eq.~(\ref{cont}) is satisfied by the {\it physical current}
$j_T (x;t) = j(x;t) + j_{\cal L} (x;t)$, which includes a contribution
$j_{\cal L}$ due to the coupling with the phonon bath. 
An explicit expression for $j_{\cal L}(x;t)$ is given in
Ref.~\onlinecite{currentOS}. In all numerical applications below we
report the total physical current $j_T(x;t)$.

\subsection{Many particle formulation}
\label{manyp}

In the many electron case, Eq.~(\ref{master}) is still valid, but 
should be reduced to an effective single-particle form in practical
applications.  This is straightforward if we adopt 
the time dependent Hartree approximation. In this approximation the
many-particle system is mapped onto a system of 
non-interacting particles,
subject to a time-dependent Hartree potential defined
in terms of the instantaneous electronic charge density \cite{TDH}.
The Hartree interaction is crucial to modeling charge rearrangement effects 
when the system is driven away from equilibrium.
 
The single-particle Hamiltonian in presence of an
electric field is $H(t) = 1/2 \left(p - 1/c A(t)\right)^2 + U(x)+ \Delta \Phi(x;t)$, 
where $\Delta \Phi$ is the periodic Hartree potential given by 
$\Delta \Phi(x;t) = \int dx' \frac{\Delta n(x';t)}{\left|x-x'\right|}$,
where $\Delta n$ is the charge density perturbation induced by the field.
 
The independent particle system is described by a 
single particle density operator $S^1$, given by $S^1 = \sum_{lm} f_{lm}
\left|l\left>\right<m\right|$ in the basis of the eigenstates of
$H_0$, {\em i.e.} the equilibrium single particle eigenstates.  
The expansion coefficients $f_{lm}$ are related
to the many particle operator $S$ via $f_{lm} = \mbox{Tr}\left[ S c^\dag_{m} c_l
\right]$. This procedure is discussed {\em e.g.} in Ref.~\onlinecite{Graham}.
The equation of motion for $f_{lm}$ then follows from Eq.~(\ref{master}). 
Using $\mbox{Tr} \left[ S c^\dag_n
  c^\dag_m c_p c_q \right] = f_{qn} f_{pm}$  (neglecting
the exchange term $-f_{qm} f_{pn}$), one finds:  
\bes
\label{eom}
\frac{d f_{nm}}{dt} &=& -i \sum_p \left( H_{np}(t) f_{pm} - f_{np}
H_{pm}(t) \right) \nonumber \\
&& + \left( \delta_{nm} - f_{nm}\right) \sum_p \left( \Gamma_{np} +
\Gamma_{mp} \right) f_{pp} \\
&& - f_{nm} \sum_p \left( \Gamma_{pn} + \Gamma_{pm}\right) \left( 1 -
f_{pp} \right). \nonumber
\ees

Here $H_{nm}(t)$  are Hamiltonian matrix elements between the equilibrium
single-particle eigenstates. In Eq.~(\ref{eom}) the diagonal elements $f_{nn}$
satisfy  $0 \le f_{nn} \le 1$ in accordance with Pauli's principle,
and the total number of electrons is conserved.
At equilibrium and in absence of an electric
field, Eq.~(\ref{eom}) gives $f_{nm} = \delta_{nm} /\left( 1 +
\exp\left( \beta (e_n - \mu) \right)\right)$, i.e. a Fermi-Dirac
distribution in which the chemical potential $\mu$ is determined by the 
condition $\sum_n f_{nn} = N$, where $N$ is the total number of electrons.
In presence of a static electric field Eq.~(\ref{eom}) leads to a stationary
distribution in which the effect of the field is balanced by the effect of inelastic
phonon scattering. In the following we find the stationary distribution 
by setting $\dot{f}=0$ in Eq.~(\ref{eom}) and solve the resulting
equation by a self-consistent procedure.

\section{Application to a DBRTS}
\label{example}

\begin{figure}
\includegraphics[angle=-90,width=\columnwidth]{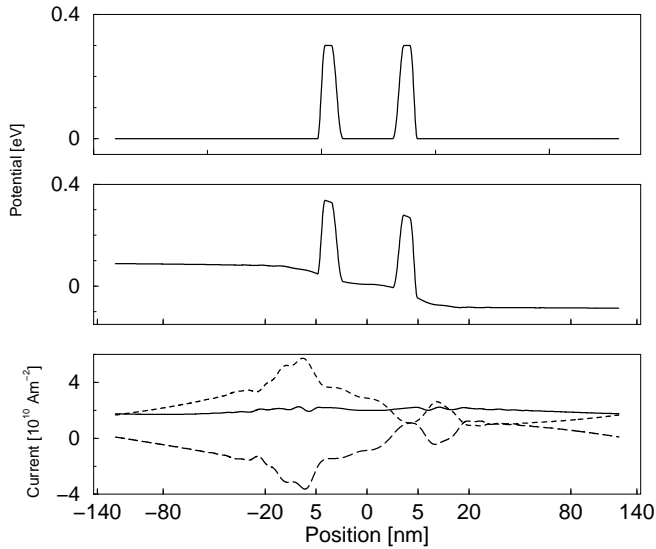}
\caption{\label{system} Upper panel: the DBRTS in the absence of an
  electric field. Middle panel: total potential in the presence of 
  external electric field for low $\gamma_0$ (see text). 
  Lower panel: spatial distribution of the current. Short
  dashed line: Hamiltonian current $j(x)$. Long dashed line:
  Current due to inelastic collisions $j_{\cal L}(x)$. Solid line: Total current
  $j_T(x) = j(x) + j_{\cal L}(x)$. Within the numerical precision of the calculation
  $j_T$ is constant as required by continuity (\ref{cont}). All plots
  use a non-linear scale for the position to magnify the barrier region.}
\end{figure}

To illustrate how this works in practice we consider a DBRTS as a test system. The
potential $U(x)$ in absence of an applied electric field is shown in
the upper panel of Fig.~\ref{system}. Outside the barrier region, we
assume a carrier density of $4.3 \cdot 10^{18} cm^{-3}$, and choose an
effective mass of $0.1 m_e$ and a dielectric constant of $10$. Due to
the large dimensions of the DBRTS in the directions perpendicular to the
barriers, an effective one-dimensional model can be assumed.
We adopt a model in which the phonon density of states ${\cal D}(\omega) \propto \omega^2$, 
and the electron-phonon
couplings $\gamma_{nm} = \gamma_0$  for all states $n,m$. We
choose values of $0.136 meV$ and of $0.218 meV$, respectively,  for $\gamma_0$, 
and a temperature of $T = 25.3 K$, such that $k_B T$ is 
comparable to the electronic level spacing 
at the Fermi energy in our finite system ($L=244 \, nm$). 
We solve Eq.~(\ref{eom}) for
steady state behavior, {\it i.e.} $\frac{d f_{nm}}{dt} = 0$, and determine 
the Hartree potential $\Delta \Phi(x)$ self-consistently, for
different values of the applied 
external field $\Efield$. The total potential acting on the electrons
when $\Efield$ corresponds to peak current is shown in the middle panel
of Fig.~\ref{system}. The total
potential includes $U(x)$, the externally applied bias (represented by a
linearly varying potential $\phi_{\Efield}(x) = -\Efield \cdot x$),
and the Hartree potential. Notice that we adopt here, for visualization
purposes, the position gauge ({\it x-gauge}) which is obviously incompatible with 
periodic boundary conditions.   
As we can 
see in the figure, the total potential has the same qualitative
behavior found in self-consistent calculations using open boundary
conditions for similar model systems. In our approach, however, the
voltage drop across the barrier is an output rather than an input
of the calculation.
Another  distinctive feature is that we observe not only a potential 
drop at the tunneling structure but also a small voltage drop along
the wire. The latter is very small (about
5 \% of the total in the case shown in
Fig.~\ref{system}), indicating that the external field is nearly
completely screened inside the wire, as one should expect.  
In the lower panel of Fig.~\ref{system} we report the current densities $j(x),j_{\cal
L}(x),$ and $j_T(x)$  corresponding to the electric field and electron-phonon coupling
of the middle panel. 
We calculate the current $j_{\cal L}$ by extending
the single-particle theory of Ref.~\onlinecite{currentOS} to the many-electron case within the time
dependent Hartree approximation. On the time scale of kinetic
evolution the current $j$ originates from Hamiltonian propagation, while the current $j_{\cal L}$ 
accounts for the charge flow due to inelastic phonon scattering 
and is larger at the contacts where deviations from equilibrium are larger. The effect is
stronger at the injection contact where charge accumulates. The behavior of $j_{\cal L}$ indicates 
where local heating should be expected to occur.

\begin{figure}
\includegraphics[angle=-90,width=\columnwidth]{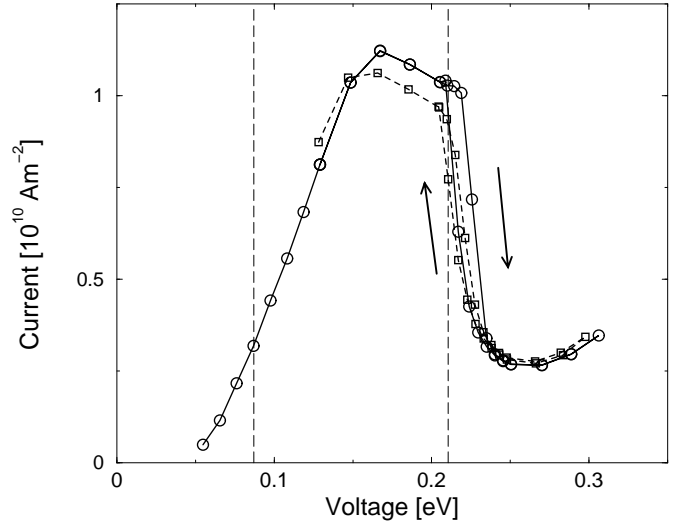}
\caption{\label{IV} Two calculated IV characteristics of a DBRTS. The curve with the
highest peak current (solid line, circles) is the one corresponding to  the lower value of $\gamma_0$.  
The energy of the dashed line on the left corresponds to twice the energy difference between 
the resonant level and the Fermi level. The second dashed vertical line
is displaced to the right by twice the energy band width. According to a simple model of resonance, 
transmission should occur in the region between the two lines.}
\end{figure}

Two $I-V$ characteristics, corresponding to two different values of the electron-phonon coupling 
parameter $\gamma_0$, are reported in Fig.~\ref{IV}. The voltage in the figure is the drop
across the barrier structure, rather than the total voltage drop $\Efield \cdot L$ in the simulation cell.  
The calculations simulate an adiabatic sweep from low to high voltage biases and viceversa:
at each small increment (decrement) of the applied electromotive force we compute a new current 
and voltage drop using the induced potential obtained in the previous step 
as input for the self-consistent calculation. Interestingly, 
the characteristics at lower $\gamma_0$ exhibits an hysteresis loop which is much less
pronounced in the 
characteristics at larger $\gamma_0$. A larger electron-phonon coupling also 
shifts the peak of the resonance to lower voltages and reduces the peak to valley ratio.   
The hysteresis loop, which is suppressed by strong electron-phonon coupling, originates
from charging of the resonant level, as already found in earlier
calculations \cite{bistability1, bistability2}.
 
Even though it is not our
aim here to reproduce a specific experiment, it is
interesting to note that by appropriate choice of the parameters 
and conditions of the calculation, we can reproduce
important experimentally observed features of a DBRTS like
bistability and hysteresis effects \cite{Tsui}. 

\begin{figure}
\includegraphics[angle=-90,width=\columnwidth]{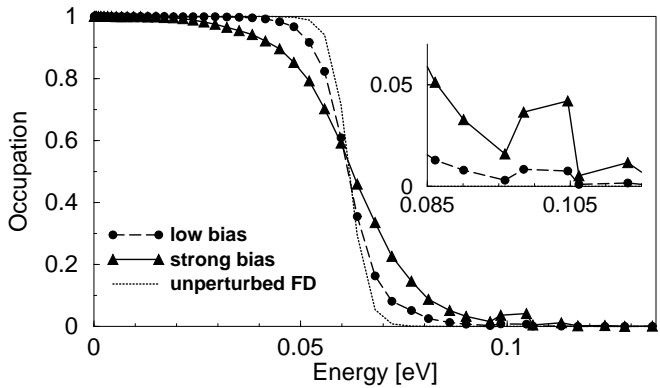}
\caption{\label{occupations} The diagonal elements $f_{nn}$ of the
  density matrix in the steady state. Stronger bias leads to a
  stronger deviation from the unperturbed Fermi-Dirac (FD) distribution.}
\end{figure}

In our approach inelastic collisions provide a source of
non-linearity other than the tunneling structure. To better appreciate
this effect we report in Fig.~\ref{occupations} the energy
distribution of $f_{nn}$, {\it i.e.} the diagonal elements of the density matrix,
for two values of $V$, a low voltage below peak current and a high voltage close to 
peak current, respectively.   
We see that at low voltage the
population is close to the thermal distribution at temperature $T$,
whereas at high voltage the population resembles a thermal
distribution at a temperature higher than $T$. This is a real
physical effect: at larger biases the electron distribution
deviates more from equilibrium. We also notice (see the inset of
Fig.~\ref{occupations}) that the
most significant deviation from thermal equilibrium is a small peak in
the tail of the distribution, in correspondence with the energy of the
resonant state. This indicates charging of the resonant level, an effect which is at 
the origin of the instrinsic bistability of a
DBRTS \cite{JAP}.

The effect of inelastic collisions on transport properties can be neglected 
in the weak bias limit (linear response regime) \cite{KohnLutt}.
In this regime transport in mesoscopic and nanoscopic systems is governed
by Landauer's formulas for the conductance \cite{Landauer}, which
can be derived within the Kubo formalism \cite{Thouless}. 
This is usually done by adopting open boundary conditions, but recently a derivation
based on a ring geometry (in the limit
$L \rightarrow \infty$) has also been reported\cite{KamenevKohn}. In the linear response regime 
the conductance can be calculated by applying a sinusoidal field 
and taking the limit $\omega \rightarrow 0$, in absence of coupling to the lattice. 
In our kinetic approach we look instead for the steady state solution under a constant 
applied field. This requires coupling to the lattice whenever $\Efield$ is
finite, and allows us to consider also strongly nonlinear regimes like
in the DBRTS case discussed above.

\section{Conclusion}

We have presented a kinetic approach to study quantum
transport at the nanoscale, including the effect of inelastic
collisions. The main physical approximations are a perturbation
treatment of electron-phonon scattering and a Markov approximation in
the dynamics. Use of simplified Hamiltonians, as in the example
discussed above, is by no means necessary. A realistic  description of
the electronic structure within Time Dependent Current-Density
Functional Theory \cite{TDDFT, TDCDFT, Kieron} is possible, and work to
achieve this goal is at an advanced stage of development. 
More accurate models for the phonons and the electron-phonon coupling would
be considered in the future. 
Our approach is not limited to 
steady-state phenomena, but could be applied 
to time-dependent transient behavior as well. 

\begin{acknowledgments}  
This work was supported by DOE under grant DE-FG02-01ER45928 and by the
National Science Foundation (MRSEC Program) through the Princeton Center
for Complex Materials (DMR 0213706).
\end{acknowledgments}

\end{document}